# Cell adhesion and fluid flow jointly initiate genotype spatial distribution in biofilms

Short title: Cell adhesion and fluid flow in biofilm early development

Ricardo Martínez-García[1], Carey D. Nadell[2,3*], Raimo Hartmann[2], Knut Drescher[2], Juan A. Bonachela[4,5*,]

1. Department of Ecology and Evolutionary Biology, Princeton University. Princeton NJ, 08544, USA. 2. Max Planck Institute for Terrestrial Microbiology, 35043 Marburg, Germany. 3. Department of Biological Sciences, Dartmouth College, Hanover, NH 03755, USA. 4. Marine Population Modeling Group, Department of Mathematics and Statistics, University of Strathclyde, Glasgow, G1 1XH, Scotland, UK. 5. Department of Ecology, Evolution, and Natural Resources, Rutgers University, 14 College Farm Road, New Brunswick, NJ 08901, USA.

*Correspondence: juan.bonachela@rutgers.edu, carey.d.nadell@Dartmouth.edu

## Abstract

Biofilms are microbial collectives that occupy a diverse array of surfaces. The function and evolution of biofilms are strongly influenced by the spatial arrangement of different strains and species within them, but how spatiotemporal distributions of different genotypes in biofilm populations originate is still underexplored. Here, we study the origins of biofilm genetic structure by combining model development, numerical simulations, and microfluidic experiments using the human pathogen *Vibrio cholerae*. Using spatial correlation functions to quantify the differences between emergent cell lineage segregation patterns, we find that strong adhesion often, but not always, maximizes the size of clonal cell clusters on flat surfaces. Counterintuitively, our model predicts that, under some conditions, investing in adhesion can reduce rather than increase clonal group size. Our results emphasize that a complex interaction of fluid flow and cell adhesiveness can underlie emergent patterns of biofilm genetic structure. This structure, in turn, has an outsize influence on how biofilm-dwelling populations function and evolve.

## Author summary

Biofilms are bacterial groups, often attached to surfaces, in which a broad variety of cooperative and competitive interactions typically occur. The spatial organization of different strains and species within biofilm communities strongly influences their global functioning, but little is known about how such structure arises. Combining experiments on *V. cholerae* and simulations of a cellular automaton, we show that the complex interaction between bacterial traits (cell adhesion) and environmental factors (fluid flow intensity) strongly influences the early origins of biofilm spatial structure. In most cases, we found that highly-adhesive strains form larger clusters than the weakly-adhesive ones. Against intuition, however, we also found the opposite outcome: weakly-adhesive tend to form larger clusters than the highly adhesive ones when flows are weak or the population density of colonizing cells is high.

# Introduction

In addition to living as planktonic cells in liquid environments, bacteria often form dense conglomerates attached to surfaces, termed biofilms. Biofilms are one of the most widespread forms of life on Earth, and they are deeply embedded into global scale processes such as biogeochemical cycling [1]. They also play a central role in the interaction between bacteria and multicellular organisms, including humans, as biofilm production enhances antibiotic tolerance [2] and influences bacterial pathogenesis and microbiome functioning [3]. From a biotechnological point of view, biofilms are used to purify wastewater and to control catalysis reactions, including those involved with biofuels [4]. Biofilms are also the primary source of biological fouling in industrial settings [5].

Within a biofilm, cells are typically embedded in a matrix of extracellular polymeric substances (EPS) made of proteins, lipids, nucleic acids and polysaccharides [6]–[8]. The secretion of the matrix, together with other products such as digestive enzymes, nutrient chelators, and adhesins, provides biofilm-dwelling bacteria with increased metabolic versatility, tolerance to exogenous stress and resistance to fluid shear [9]–[15]. The functioning and evolutionary stability of behaviors that alter the local environment – including secretion phenotypes, which usually affect nearest-neighbors the most strongly – in turn depend on the spatial arrangement of secreting versus non-secreting strains and species (i.e., different genotypes) in a biofilm community [16]. For example, intra-strain cooperative behaviors are more likely to be evolutionarily stable when different cell lineages are segregated in space, with typical interaction distances between cells being strongly influenced by the diffusivity of secreted products, biofilm architecture, and environmental flow conditions [16]–[19]. Spatially constrained interaction is well known to be important in ecology broadly, and there are numerous examples of spatial structure influencing evolution in biofilm communities [20]–[22]. Thus, spatial structure in biofilms, once it arises, has a large impact on their form and function. The means by which biofilm strain and species structure originates in the first place, however, are less well understood.

At the early stages of biofilm formation, planktonic cells encounter and transiently adhere to surfaces. Bacteria possess sophisticated mechanisms for deciding whether to remain in place, depending on substratum properties and environmental quality [23]–[26]. Having committed to biofilm formation, surface-residing cells secrete additional and diverse adhesion factors, including extracellular matrix material. These secretions, in combination with growth, death, and steric interactions between cells, strongly impact biofilm spatial organization [16], [27]–[30]. Environmental features, such as surface chemistry and fluid flow, are also key to biofilm development. In cases where flow influences cell surface motility, flow regime and environmental geometry can exert a dramatic effect on the spatial spread of surface-bound bacteria [31], [32]. Fluid flow is also likely to play a key role in the deposition and spatial arrangement of different strains and species within biofilms [15], [33]–[35]. In spite of its putative importance, we have a limited understanding of how flow, surface colonization processes, and cell adhesion interact to influence the spatial strain structure of nascent bacterial communities. Targeting this knowledge gap is the primary goal of the present study.

We performed experiments with matrix-producing or non-producing strains of the model biofilm-forming bacterium *Vibrio cholerae*, the causative agent of pandemic cholera in humans. We aimed to use a simplified, ecologically neutral scenario, in which mixed strains are genetically identical except for fluorescent labels, to provide a first step towards understanding how key environmental features interact with cell adhesion and population density to control the initial distribution of cell lineages on a surface [36], [37]. Based on these experiments, we developed a cellular automaton, with which we considered different scenarios that included varying flow strengths, densities of founder cells, and variable cell adhesiveness. Our study of surface occupation patterns motivated the use of spatial correlation functions as a quantitative method to characterize the contribution of adhesiveness and flow regime on the origins of clonal clustering spatial structure. The results, although obtained for *V. cholerae,* will more generally improve our understanding of the patterns with which microbes colonize abiotic and biotic surfaces. These initial patterns of surface occupation are key to the longer-term biofilm architectures that endure to impact bacterial ecology, evolution, and pathogenesis.

## Results

### Surface colonization experiments

To isolate the influences of adhesiveness, flow, and population density on surface colonization regimes, we used strains of *V. cholerae* without flagella that either produce extracellular matrix constitutively, or not at all [38]. As *V. cholerae* does not use gliding or twitching motility to roam on glass surfaces after attachment [39], differences in surface occupation by our strains could be specifically attributed to their difference in matrix production. Individual cells of *Vibrio cholerae* are capable of attaching to surfaces in the absence of extracellular matrix secretion, but matrix production augments surface and cell-cell adhesion, and is essential for producing three-dimensional biofilm structures. The direct contribution of matrix production to biomass accumulation in biofilms, relative to loss of cells into the passing flow, has been demonstrated in our previous work [38], [40]. Cell motility in the planktonic phase, which influences surface exploration [39], [41]–[45], is not included here and will be the focus of future work.

For each strain (matrix-producing, and non-producing), a red- and blue-fluorescent version was constructed by engineering fluorescent protein expression constructs on the chromosome. Founder cells were inoculated in polydimethylsiloxane (PDMS) microfluidic chambers as 1:1 co-cultures of the blue and red variants of the matrix-producing strain, or, in separate experiments, blue and red variants of the matrix non-producing strain. Flow rate was maintained at 0.1 µL/min through chambers measuring 500 µm wide, 100 µm tall, and 7000 µm long. Bacteria within a growth chamber were thus identical regarding the production of matrix, differing only in their color. Nutrients were continuously provided in the inflow. We focused on the early stages of biofilm growth before large 3D structures could form; thus, growth was limited by the availability of space on the surface, and not by the access to nutrients in the influent medium. It is important to note that, even in these early phases of biofilm growth (once cell clusters reach 4-8 bacteria), cells capable of producing matrix will have begun to do so [46].

Experiments were stopped when the biofilm population fully occupied the basal surface, as judged by eye. The data generated by our experiments consisted of 2D surface occupation patterns composed by clusters of different lineages that express either the blue or the red fluorescent tag (Fig 1). Surface occupation was captured by fluorescence microscopy. Images were acquired in the largest viewing fields allowed by our microscope constraints, measuring 60 µm x 60 µm (923 x 923 pixels), with 60 such viewing fields comprising an entire chamber. Note that, since the snapshots analyzed in the experiments correspond to tiles within a larger total area in the growth channel, there can be exchange of cells across tiles through detachment and re-attachment of individuals. See Materials and Methods for a more detailed description of our experimental approaches and strain engineering, and S1 Fig.a for a schematic representation of the experimental procedure.

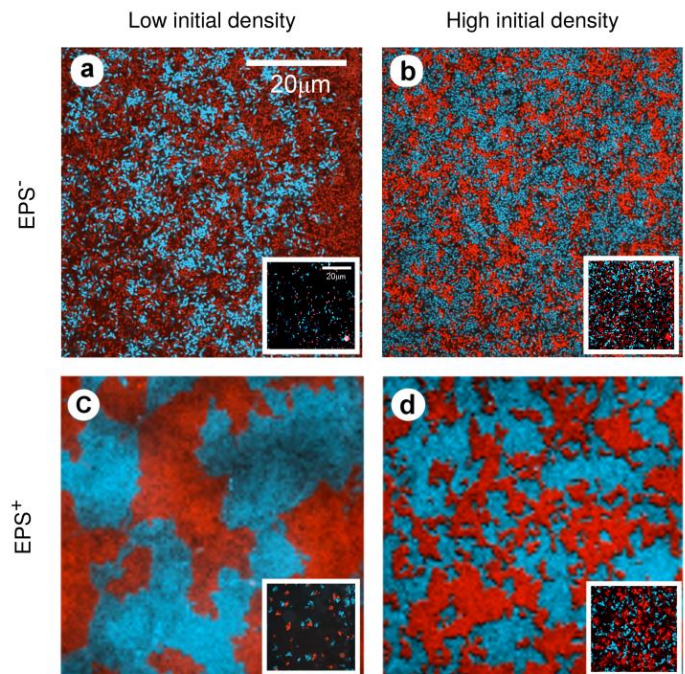

**Fig. 1. Experimental colonization patterns.** Snapshots of one field of view at confluence for both matrix-producing (bottom) and non-producing (top) strains at low (left) and high (right) initial cell densities. The inset of each panel shows the initial distribution of founder cells. Initial densities: a) 0.01 cells/µm², b) 0.162 cells/µm², c) 0.012 cells/µm², d) 0.113 cells/µm².

## Modelling framework

To explore the mechanisms underlying the experimental results, and to extend our predictions to a broader set of environmental flow conditions and cell adhesion strengths, we developed a probabilistic cellular automaton capturing the essential features of the experimental system. In our model, we consider two strains with identical non-dimensional cell adhesiveness, $\sigma$, and initial density of colonizing cells, $\rho_0/2$, that compete for the occupation of empty space on a discrete two-dimensional lattice. The density of founder cells is defined by the fraction of initially occupied lattice squares. In the absence of extensive surface motility, adhesiveness varies inversely with the probability that a cell detaches from the surface. This may occur either because of shoving between cells or because of flow, which detaches cells and relocates them downstream. We will use here a real number in [0,1] to represent adhesiveness, with $\sigma = 1$ indicating strong adhesion and $\sigma = 0$ weak adhesion. The only difference between strains within a given experiment is therefore a binary variable for the cell color, $c$, which is later used to analyze the arrangement of different cell lineages.

The dynamics of the model has two main ingredients: (*i*) birth and (*ii*) flow-induced cell detachment and relocation. We assume that these two processes are stochastic and independent (S1 Fig b,c). Time is discretized in short intervals of fixed length $dt$; within each time step, a random cell reproduces (*i.e.* divides) with probability $p_b$, and another random cell may be detached and eventually relocated with probability $p_d$. The detachment probability depends on cell adhesiveness and flow intensity, whereas cell transport, both in the direction of the flow and transversely to it, is entirely determined by flow intensity, $f$, which we define using a normalized non-dimensional parameter that takes values in [0,1]. The flow structure in our microfluidic devices is laminar, so we assume that flow intensity fixes the maximum distance that cells may be transported downstream. $f = 1$ represents intense flows under which cells can be transported a maximum distance equal to lattice length, and $f = 0$ represents no flow and therefore no cell detachment and transport. Cell transport in the direction transverse to the flow is bounded by the distance traveled downstream (see Materials and Methods). Since surface colonization occurs over short time scales and resources are continuously supplied by the inflowing nutrient medium, we do not include cell death in the model. In our experiments cells can in principle detach from one viewing field and re-attach in another viewing field downstream; we implement this possibility in our simulations using periodic boundary conditions. Cells that exit the system through one of the borders due to long-range relocation re-enter through the opposite side, which is equivalent to cell relocations originating upstream and balances out the anisotropic effects introduced by the presence of a directional flow.

Finally, each run of the model was stopped when 95% of the positions of the lattice were occupied, which avoids the high number of shoves that occur when surface coverage is nearly complete and which have a negligible effect on the final coverage pattern. This condition is similar to that used in terminating the experimental runs, which were stopped when the bottom surface of the chamber was nearly completely covered by cells. See Materials and Methods for further details on the modeling approach.

**Experimental output and model validation.**

To characterize the patterns of bacterial surface occupation obtained experimentally (Fig 1), we measured their clonal correlation lengths, $\xi$, and studied their dependence on the initial population density. The correlation length is obtained from the spatial autocorrelation function, $C(r)$, which provides a measure of the order in spatially-extended systems by quantifying how its spatial elements co-vary with one another on average, as a function of spatial separation distance $r$. For a given separation distance $r$, the autocorrelation is positive if individuals separated by $r$ tend to be of the same type, negative if they tend to be of different types, and zero if there is no consistent relationship between them. The correlation length is, thus, the shortest distance for which two spatial elements of the patterns are statistically independent. See Materials and Methods for a detailed definition of the correlation function. Because this distance is related to the typical cluster size within the field-of-view, from an ecological perspective, the mean correlation length quantifies the expected lineage segregation in the surface occupation pattern (see Materials and Methods). When two matrix-secreting strains colonize the chamber, the correlation length of the confluence pattern increases as the total initial density of cells decreases (green dots in Fig 2). However, if the two strains are matrix non-secreting (and therefore only very weakly adhesive), the correlation length does not show strong dependence on the initial density of cells in the chamber (black dots in Fig 2). Note that the lowest initial coverage densities for the two cases are different; matrix-secretors could be initiated at very low densities for which non-secreting strains did not give viable results. This limitation on initial population density was most likely due to the relative ease with which non-secreting strains are removed by flow.

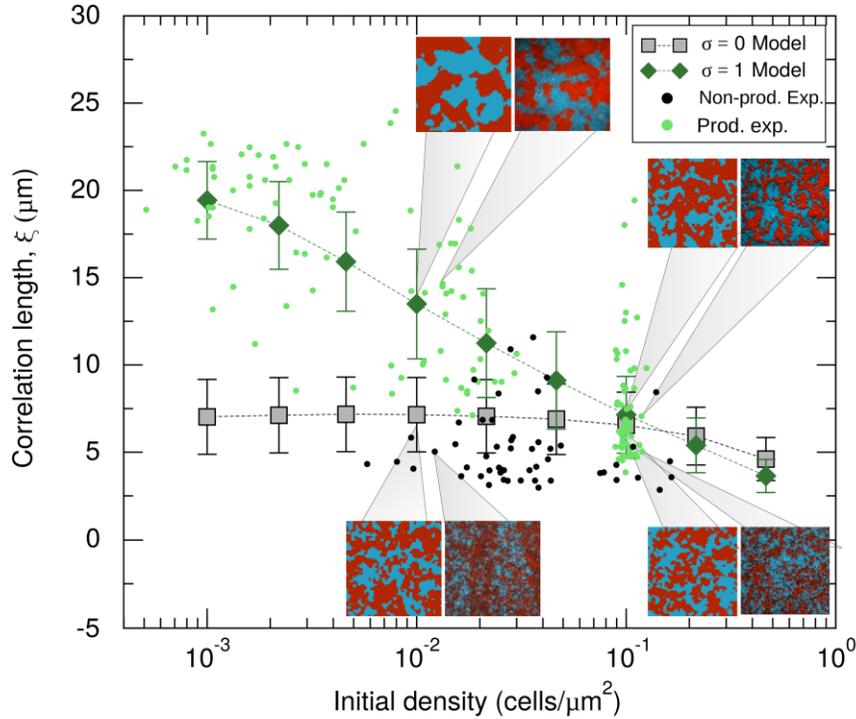

**Fig 2. Model validation: correlation length comparison**. Experimental correlation lengths measured in the matrix-producing (pale green dots) and non-producing (black dots) strain, and their model equivalent $\sigma = 1$ (dark-green diamonds), respectively $\sigma = 0$ (gray squares). Numerical results are shown for flow intensity $f = 1$, which gives the best agreement with the experiments, averages taken over $2 \times 10^6$ independent realizations. Error bars represent the standard deviation. The insets show snapshots of colonization patterns obtained in the experiments (right) and the model (left) at initial colonization densities indicated by the gray pointers.

To compare our model and experiments, we used the simulation framework to study the behavior of the clonal correlation length as a function of flow intensity and system size. To keep our analysis as close as possible to the experiments, we initialized each simulation with a density of cells $\rho_0$ and assigned to each cell either the blue or red color with probability 0.5. In this manner, we constructed, on average, a 1:1 (blue:red) mix of cells randomly located within the lattice. Since bacteria in our experiments either produce matrix constitutively or not at all, we assumed that these strains correspond in our model to the $\sigma = 1$ (highly-adhesive) and $\sigma = 0$ (weakly-adhesive) cases, respectively. In addition, we parametrized the spatial scale of the model to mimic the experimental device. We used a square lattice of lateral length $L = 60$ sites, which represents each of the (60 μm x 60 μm) field-of-view tiles of the experimental system (i.e., corresponding to a lattice mesh size $dx = dy = 1$ μm), and assuming an approximate cellular cross section of 1 μm² [29], we limited the maximum occupancy of each position of the lattice to only one cell. Finally, since we are interested in the final occupancy patterns, regardless of the temporal scale at which colonization takes place, we fixed the birth rate to minimize the computational time. This parametrization leaves flow intensity, $f$, as the only parameter that is free in the model but fixed in the experiments. Since flow intensity is defined in terms of a non-dimensional quantity in the model, we established a connection between its value in the experiments and the model by finding the best quantitative agreement between model-produced and experimental patterns. For a broad range of flow intensities (S2 Fig), the theoretical results confirm the qualitative trend observed with our experiments: clonal correlation length and total initial density are negatively correlated for highly-adhesive cells, but nearly uncorrelated for weakly-adhesive cells. However, we found the best quantitative agreement for the mean correlation length between experiments and simulations in the strong flow limit $f = 1$, for which the simulation results are shown together with the experimental data in Fig 2. The correlation length is also quantitatively, but not qualitatively, affected by the simulated "field of view" (or tile size); spatial segregation increases for larger systems, but the trends of the $\sigma = 0$ and $\sigma = 1$ curves are independent of system size for $f = 1$. A more detailed analysis of the effect of system size in our simulations is provided in S1 Text.

As a last part of the model validation effort, we obtained the simulation (highly-adhesive, $\sigma = 1$) and experimental (matrix-producer) distributions resulting from the correlation lengths obtained with independent replicates, and compared one versus the other for different initial densities (Fig 3). To compute the distributions, we divided the experimental measures in three ranges of initial densities (low, intermediate and high according to the clusters of experimental data observed in Fig 2) and used fast adaptive kernel density estimation in which the bandwidth of the kernel varies across the dataset. These algorithms are particularly useful to estimate asymmetric distributions with a fat tail in one extreme and a thinner tail on the other [47]. The model and experimental distributions agree, especially in the high and low-density limits at which more experimental replicates were gathered. Note that, in both extremes, the estimated distributions are skewed (S3 Fig), suggesting that the median is a better measure of the central tendency of the distribution than the mean. However, because both measures do not seem to differ significantly (see S2 Fig and S4 Fig) whereas the mean provides less noisy results, we will focus hereafter on the mean and the standard deviation as indicators of central tendency and dispersion, respectively.

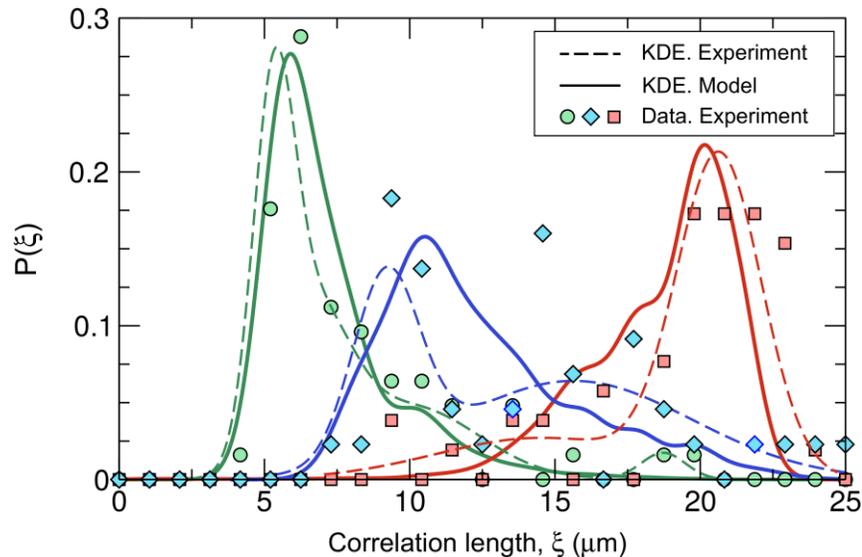

**Fig 3. Model validation: distribution of correlation lengths for matrix-producing strains.** Estimated theoretical (full line) and experimental (dashed line) correlation length distributions. The symbols represent the experimental distribution prior to smoothing estimations. Each color represents a range of colonizing cell densities: green, $10^{-1}$ cells/$\mu m^2$ for the model and high

density experimental data (cluster of data around $10^{-1}$ cells/µm$^2$ in Fig 2); blue, $10^{-2}$ and $2.15\text{x}10^{-2}$ cells/µm$^2$ for the model and intermediate density experimental data ($7\text{x}10^{-3} < \rho_0 < 3\text{x}10^{-2}$ cells/µm$^2$); and red, $10^{-3}$ and $2.15\text{x}10^{-3}$ cells/µm$^2$ in the model and low density experimental data ($\rho_0 < 5\text{x}10^{-3}$ cells/µm$^2$).

## Model predictions. Interaction between bacterial traits and flow intensity.

As discussed above, we consider founder density and adhesiveness as the traits of interest for our strains in this study. Both of these traits are influenced by genetically encoded factors, such as matrix secretion, as well as by environmental factors, such as habitat turnover and surface chemistry [48]. To the extent that adhesion and surface colonization density are under bacterial control, we consider these traits here to be part of a general strategy set for influencing surface occupation [49]. We explore the effects of the flow on colonization strategies by studying how diverse combinations of flow strength, adhesiveness, and initial population density influence final patterns of surface occupation. As described above, numerical simulations were initiated with a 1:1 mixture of red and blue strains that have the same adhesiveness.

As shown in S2 Fig, the mean correlation length decreases as the initial density increases for any flow intensity and any cell adhesiveness. This trend is applicable also for weakly adhesive cells ($\sigma = 0$), although for the highest flow intensities the trend is only evident for very high initial densities. The results are more convoluted when looking at a range of adhesiveness for a fixed initial density (S5 Fig). Lower $\rho_0$ (cells/µm$^2$) conditions show null or positive association between adhesiveness and correlation length, whereas higher initial densities show a null or slightly negative interdependence.

In order to assess how the different colonization strategies would be influenced by the flow, we quantified the difference between the correlation length reached by highly-adhesive strains ($\sigma = 1$) and weakly-adhesive strains ($\sigma = 0$) as a function of flow intensity and initial population density. Intuitively, one might expect that populations of highly-adhesive cells universally obtain larger clonal clusters, and indeed, this outcome does occur broadly, especially with increasing flow speed. When flow is strong, less-adhesive cells are frequently removed from the surface, exposing new area for attachment and growth and generally causing population admixture. However, there is a considerable region of the parameter space in which populations of weakly-adhesive cells show the larger clonal clusters (higher correlation length) at confluence, especially when flow is weak, or when initial population density is high (Fig 4a). The difference in correlation length between highly and weakly-adhesive cells becomes more pronounced in larger systems. Weakly-adhesive strains form larger clusters than the highly-adhesive ones for a larger set of flow intensities, and this difference in cluster size can be quantitatively of similar magnitude to the one gained by highly-adhesive strains in the strong flow limit (further details on the effects of the system size are provided in S1 Text).

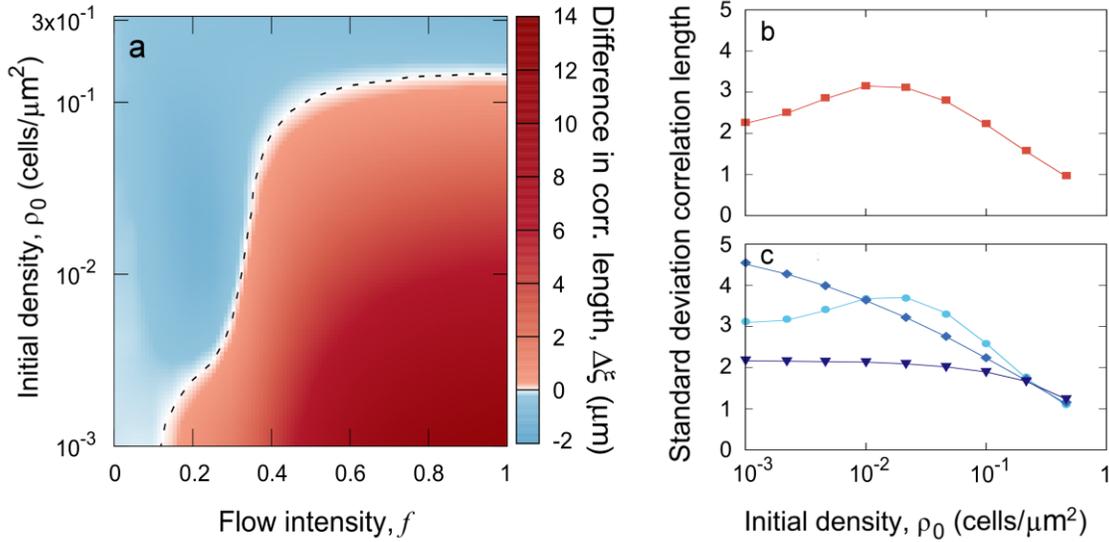

**Fig 4. Model output: mean cluster size and variability.** a) Difference in correlation length resulting from investing in cell adhesion for different flow intensities and initial colonization densities. The dashed line indicates the values of $f$ and $\rho_0$ at which this difference is equal to zero. Averages are taken over $5 \times 10^4$ independent realizations. b, c) The standard deviation of the correlation length is a proxy for lineage segregation variability in highly-adhesive strains, (b; $\sigma = 1$) and weakly-adhesive cells (c; $\sigma = 0$). Averages are taken over $2 \times 10^6$ independent realizations of the model.

Finally, the correlation length of clonal clusters is highly variable in our experiments with constitutively matrix-secreting cells, especially for intermediate colonizing population density (Fig 3). In light of this observation, we used the model to investigate how flow intensity influences variability in the correlation length for highly- and weakly-adhesive cells and continuously varying initial surface density. For low flows, the variability in clonal cluster size follows the same trend for highly-adhesive and weakly-adhesive strains, reaching its maximum values at intermediate initial densities (Fig 4b, 4c, S6 Fig). Differences between strains emerge as flow intensity increases. On the one hand, highly-adhesive cells cannot be detached or shoved, and thus their cluster size variability is not influenced by flow speed (S6 Fig). Such variability in the correlation length is also quantitatively influenced by system size, although the curve maintains its concavity as a function of the initial population density (S1 Text). On the other hand, as flow speed increases, the dispersion in the weakly-adhesive strain correlation length transitions from a convex form to a uniformly decreasing function of initial population density (Fig 4c). This pattern holds for strains with intermediate adhesiveness, although the influence of flow intensity on correlation length variability decreases as adhesiveness increases (S6 Fig).

## Discussion

Combining experiments in microfluidic devices with numerical simulations of a cellular automaton, we have developed a framework for quantifying strain mixture versus segregation in the coverage patterns that emerge from bacterial expansion competition on 2-D flat surfaces. We used experimental data to validate the core assumptions of the model framework, which permitted us to make predictions for a broad set of ecological scenarios defined by the intensity of environmental flow, surface colonization density, cell adhesion properties, and the extension of the colonized surface.

Microbes occupy a vast variety of surfaces, often subject to a wide range of fluid flow intensities. A common example of surface attachment stressed by laminar flow-induced shear forces is chitin colonization in marine environments [9], an important feature of the natural ecology of many *Vibrio* species. Typical surface colonization densities are also likely to vary widely depending on the species, environmental conditions, and local demographics of bacterial communities. Among the mechanisms that control seeding density, some are under bacterial control, and others are not. For example, chemotaxis toward surfaces and the active production of adhesins/extracellular matrix can modulate cell surface occupation, but so too will ambient population density conditions in the planktonic phase, local flow patterns, and the chemistry of the surface bacteria attempt to colonize [25], [48], [50], [51]. Decreasing the initial colonization density increases the typical distance between founder cells and thus the territory that can be potentially occupied by each of them and its descendants [52]. In our experiments and simulations with highly-adhesive strains subject to strong flows, this translates into a negative correlation between cell lineage cluster size and initial cell density, consistent with previous reports in other species [53]. In populations of weakly-adhesive cells, however, flow encourages spatial mixing of genetic lineages by detaching cells and transporting them to other positions in the local environment, which reduces the sensitivity of the final pattern to the initial conditions. As a result, when flows are strong and colonization densities are moderate to low, investment in cell-cell and cell-surface adhesion results in stronger clonal clustering of cell lineages.

It follows from intuition that populations of highly-adhesive cells might generate coherent clonal clusters more easily than less adhesive cells. And indeed, this result was observed in our experiments and for many model conditions. However, there was a broad region of the model parameter space in which the opposite behavior was predicted. This exception occurred at low flow strengths and, independently of flow strength, when the initial population density of colonizing cells was very high. In each of these two cases, we found that a different mechanism underlies such counterintuitive result. For the former case, if flows are weak cell relocations occur over short distances, which alleviates local competition for space within large clusters instead of mixing the population. Weakly adhesive strains thus form larger clusters than highly-adhesive strains *via* limited dispersal. For the latter case, when surfaces are almost fully occupied during the colonization phase, populations of highly adherent cells (which resist removal by flow) fix the initial state of the system into one of randomly distributed cell lineages. In populations of weakly adhesive cells, however, the vast majority of cells that detach cannot re-attach to the surface elsewhere and are lost to the flow output. The positions from which detached cells were removed are then occupied by descendants of neighboring cells that had managed to remain in place. If the detached cell was originally surrounded by cells of its same lineage, then the empty space is filled by a new cell within the same lineage and the update has no effect; in a mixed region, however, the growth will tend to reduce mixing and thus to increase the clonal correlation length of the system. Therefore, populations of highly-adhesive cells are not universally expected to show stronger spatial genetic structure than populations of less adherent cells; the structure depends on the ecological conditions and bacterial traits controlling surface colonization density, as well as the environmental flow regime.

Complex surface attributes, such as its topology and chemical properties, are not explored here but are expected to influence cluster sizes in some natural environments by increasing the complexity of fluid flow patterns, inducing short-range cell relocation and modifying the long-range relocation mechanism. Furthermore, in our simulations and experiments, surfaces are unoccupied prior to cell inoculation. In *V. cholerae* and other biofilm-forming organisms, matrix production is known to prevent planktonic cells from entering the biofilm, thus providing a competitive advantage to resident cells during surface colonization processes [54]. The tendency of cells to adhere to one another and form large clusters is likely to fall under selection based on the size of resource patches in a given environment. Resources matching has been intensively addressed in animal ecology, both from the perspective of optimizing the search process [55]–[58], and including its demographic implications [59]–[61]. Given our model results, we speculate a relationship between adhesion, surface attraction, and the variance of cell lineage cluster size that could determine the ability of microbes to cope with variability in nutrient patchiness. Exploring the role of these three parameters is a future line of research expanding upon this study.

The emergent spatial structure of cell lineages during biofilm growth is important to numerous other facets of microbial ecology, especially for the evolutionary trajectories of social phenotypes [16]. Many phenotypes associated with biofilm formation and the pathogenesis of bacterial infections, for example, are secreted factors such as digestive enzymes and nutrient-chelating molecules [66]. In many cases, these secreted compounds may enable a biofilm, as a collective, to degrade complex polymers – including host tissues – that otherwise would be inaccessible [9], [12]. Since secreted enzymes can be costly to produce and may benefit all cells in the immediate surroundings, their evolutionary stability often relies on population structure, which can promote preferential interaction among cells of a single strain. If cells are mostly surrounded by neighbors of the same lineage, cooperative cells are more likely to interact with clonemates, which are also cooperative, promoting the evolutionary stability of the cooperative phenotype in question [21], [67]. Other forms of cell-cell interaction, on the other hand, are only effective in mixed population structures; these include, for example, cross-feeding mutualisms in which different cell types depend on close proximity to benefit each other [17]–[19]. Antagonistic phenotypes, such as toxin secretion (e.g. Type VI-mediated attack), also depend on mixed population structure to be optimally effective [68]–[73].

Given the relationship between spatial structure and the evolutionary stability of different secretion phenotypes, we might expect surface colonization and adhesion strategies to coevolve with the ability to produce extracellular public goods, as well as toxins. This would be consistent with the coevolution of cooperation and dispersal more generally, either via movement in motile organisms or passive transport in sessile species, which has been well-explored [74]–[80]. Varying surface colonization and adhesion is just one of several means through which spatial structure can be altered by microbes in the process of biofilm growth [16]. Previous reports have shown that some organisms, such as the social amoeba *Dictyostelium discoideum*, preferentially adhere to clonemates and promote aggregation of genotypes during collective movement [81]. For many microbes, the expansion of growing cell groups toward a source of limiting nutrients tends to promote the spontaneous segregation of different strains due to genetic drift along the advancing group front [36]. After colonizing a surface, matrix-guided motility heavily influences early biofilm structure in some strains of *Pseudomonas aeruginosa* [82]. During cell group growth, phenotypes like toxin secretion also promote strain segregation by enforcing positive feedback on the local frequency of each self-immune toxin-secreting strain of *V. cholerae* [72]. Combined with constraints due to surface properties, this array of biological forces can yield complex dynamics of spatial organization in microbial communities that we are just beginning to understand [22].

Other factors will also impact the evolution of adhesion phenotypes, including the relative advantage of highly-adhesive cells against less-adhesive cells in direct competition [54], and the trade-off between competitive surface adhesion and the ability to disperse to new habitats for later growth [38], [40], [83]. This is only one of many dimensions of surface-associated microbial behavior, which can include sophisticated mechanisms of surface departure and re-attachment, as well as various forms of individual and collective surface motility [26]. Disentangling the impacts of these different adhesion and detachment principles is an important area for future study.

## Materials and Methods

### *V. cholerae* strain engineering.

We conducted surface colonization experiments using *V. cholerae,* a model organism for biofilm formation on a broad range of surfaces. In order to control the several genes that are regulated by the flagellum activity and by quorum sensing, we first deleted *flaA,* which encodes the flagellin core protein, and *hapR*, which encodes the quorum sensing master regulator. This results in a double mutant *ΔflaAΔhapR* that produces EPS and therefore termed EPS$^+$. Second, we produced a triple mutant strain by deleting *vpsL*, a gene required for EPS production. The resulting *ΔflaAΔhapRΔvpsL* strain never produces EPS and we thus call it EPS$^-$. Finally, we derived two versions of the EPS$^+$ and the EPS$^-$ strain: one that expresses the teal fluorescent protein *mTFP1* and one that expresses the ref fluorescent protein *mKate*. This difference in the fluorescence protein is the only difference between otherwise genetically identical strains in our mixes, and it will allow us to distinguish different lineages in the surface colonization pattern.

### Experimental protocol.

We performed bacterial surface occupation experiments using microfluidic culture methods. Chambers were 500 μm wide, 100 μm high and 7 mm long, and were constructed from poly(dimethylsiloxane) bonded to glass coverslips. Overnight cultures of the EPS+ and EPS- strains were normalized to an optical density at 600 nm of 1.0, mixed to create a 1:1 culture of red and blue cells, and back-diluted either 1:100, 1:1000, or 1:10000 prior to being introduced into the chambers. The cultures were then incubated at room temperature for one hour to allow cells to attach to the glass coverslip. Varying the planktonic culture density in this manner allowed us to vary the initial population density on the glass surface. Following this attachment period, sterile M9 medium with 0.5% glucose was introduced to the chamber at 0.1 uL/min, using a high precision syringe pump (Harvard Apparatus). The chambers were fixed to the stage of an inverted spinning disk confocal microscope (Nikon, Andor), which was used to capture images of the cell populations residing on the coverslip glass. The entire surface of each chamber was imaged once per hour until surface coverage was complete as judged by eye.

### Model details.

The two main ingredients of our model are:

*(i) Reproduction.* Bacteria reproduce at a given rate $\mu$: every time step a cell division takes place with probability $p_b = \mu \, dt$, where the length of the time step $dt$ (i.e. temporal resolution of the simulations) is fixed such that $p_b < 1$. Since fluorescent protein constructs have no fitness effect [38], we set the same reproduction rate for both strains. In addition, since in each experiment both strains equally invest in adhesion, we ignore the potential cost of adhesiveness here. Finally, since we are interested in the final spatial cell distributions, our results are independent of the specific value used for $\mu$, which is fixed to minimize computational time. Newborn cells occupy a randomly chosen site among the available places within the Moore neighborhood of the parental cell (eight lattice positions surrounding the parental cell). If there is no empty position, the new cell will try to shove one of the resident (i.e. existing) cells in the neighborhood and occupy its position. The outcome of a shoving attempt is determined by a displacement probability, $p_s$, defined in terms of non-dimensional adhesiveness as:

$$p_s = \frac{1 - \sigma}{2}. \tag{1}$$

With this definition, highly-adhesive cells ($\sigma = 1$) are never displaced by newborns, whereas weakly-adhesive residents and newborns will have the same probability to be shoved due to low cell-surface adhesion (i.e. $p_s$=0.5). From each shoving event, two possibilities ensue: (*i*) the resident cell remains in its position, and the newborn is displaced to one of the empty neighboring sites of the resident, or (*ii*) the newborn cell takes the position of the resident, which is displaced to one of its empty neighboring sites. In both scenarios, if the complete neighborhood of the resident cell is occupied, the losing cell is removed from the system with the outflow. Note that this formulation truncates a cascade of shoving events that might take place for weakly adhesive cells as a cluster of bacteria expands from its center. In this manner, we are assuming that shoving events can only occur on short spatial scales before one cell must be released into the passing flow to relieve the pressure of increasing local population density.

(*ii*) *Cell detachment and relocation.* At every time step, we also check for potential cell relocations that occur due to fluid flow passing above the cell monolayer. Since flow enters the experimental chambers from one direction only (left to right), we assume that cells can only be removed by flow if their neighboring position on the left is empty. This simplification implements a drafting effect that is supported by basic fluid mechanics calculations reported by [84]: cells are protected from drag by neighbors that sit on the surface immediately upstream. Therefore, the detachment probability is zero when the directly adjacent up-stream site is occupied, and $p_d$ otherwise. We define $p_d$ using a combination of the non-dimensional flow strength, $f$, and cell adhesiveness:

$$p_d = f(1 - \sigma), \tag{2}$$

where, for simplicity, we assume that $f$ is normalized and therefore can take any value between 0 and 1. According to Eq. (2), highly-adhesive cells cannot be detached, whereas weakly-adhesive cells will be dislocated with a probability given only by the strength of the flow. Because it is not possible experimentally to track detached and re-attached individual cells over the full length of the microfluidic growth chambers to inform our model, we hypothesized a mechanism for long-range surface re-attachment. We could thus make predictions of the spatial structure of the population at confluence and directly check them against experimental results. In our simulations, once a cell has been detached, a landing position is calculated using the following rules that account for flow directionality. The distance traveled in the direction of the flow, $\Delta x$, is determined by a random integer uniformly distributed between 0 and $fL$, whereas the distance traveled in the transversal direction, $\Delta y$, is obtained as a random integer uniformly distributed between $-\Delta x$ and $\Delta x$. If the sorted position was already occupied, then the detached cell is removed from the system, which accounts for bacterial loss with the outflow. With these rules, cells can only relocate to positions downstream of the flow orientation, unless they pass through the system boundaries due to periodic boundary conditions, which recovers the isotropy in the surface-occupation patterns. On the other hand, detached cells can freely drift perpendicular to the flow. A summary of the model parameters and their numerical values is provided in S1 Table.

**Characterization of surface occupation patterns: the correlation length.**

We characterize bacterial surface occupation patterns using the spatial autocorrelation function, $C(r)$, which can be mathematically defined as,

$$C(r) = \frac{<c(R)c(R+r)> - <c(R)><c(R+r)>}{<c^2(R)> - <c(R)>^2} \qquad (3)$$

where $c$ is the binary variable that represents the lineage color (and thus takes value 1 or 2 depending on whether the lattice cell is occupied by a blue or red cell), and $<.>$ represents an average over all the elementary spatial units of the system, which are labeled by the index $R$. Given the use of periodic boundary conditions in our cellular automaton and cell mixing across adjacent tiles in the experimental device, surface occupation patterns are isotropic and the average over the angular variable can be done.

In microscopy images, the elementary unit is the pixel (0.065 µm), whereas in the simulations, it is the lattice position (1 µm). Note that the normalization factor ensures that the correlation function reaches 1 when two positions have a perfect correlation. In addition, the uncorrelated average product, $<c(R)><c(R+r)>$, force the correlation function to be zero when two locations are completely independent from each other. The correlation length is thus given by the first zero of the correlation function (S7 Fig). The spatial autocorrelation function given in Eq. (3) is related to the radial distribution function, often used to describe how density varies as a function of distance from a reference particle in a system of multiple particles.

## Acknowledgments

The authors would like to thank J. Dushoff for early discussions related to the project, to E. Bendo for assistance in data collection during early stages of the experiments, and to S.A. Levin for early discussions and a critical reading of the manuscript. They are also grateful to IFISC (CSIC-UIB) for the use of its computational infrastructure. This work is funded by the Gordon and Betty Moore Foundation through Grant GBMF2550.06 to R.M-G; J.A.B. is supported by the Marine Alliance for Science and Technology for Scotland (MASTS) pooling initiative, funded by the Scottish Funding Council (HR09011) and contributing institutions. C.D.N. received support from the Alexander von Humboldt Foundation and the Cystic Fibrosis Foundation (STANTO15RO). K.D. received support for this work from the Max Planck Society, Behrens Weise Foundation, European Research Council (StG-716734), Human Frontier Science Program (CDA00084/2015-C), and Deutsche Forschungsgemeinschaft (SFB987).

# Supporting information

| Symbol | Name | Cause | Value |
|--------|------|-------|-------|
| $\sigma$ | Cell adhesiveness | BT | Free parameter in [0,1] |
| $f$ | Flow intensity | EF | Free parameter in [0,1] |
| $\rho_0$ | Founder cell density | BT by EF interaction | Free parameter in [$10^{-3}$, 0.5] $cells/\mu m^2$ |
| $\mu$ | Reproduction rate | BT | Fixed parameter, 0.57 a.u. |
| $L$ | Lateral lattice size | EF | Fixed parameter, 60 $\mu m$ |
| $dt$ | Time step | -- | Fixed parameter, $1/L^2$ a.u. |
| $dx$ | Lattice mesh | -- | Fixed parameter, 1 $\mu m$ |
| $p_d$ | Detachment probability | BT by EF interaction | $f(1 - \sigma)$ |
| $p_s$ | Shoving probability | BT by BT interaction | $\dfrac{1 - \sigma}{2}$ |
| $\Delta x$ | $x$-distance traveled | EF | Random, in [0, $fL$] |
| $\Delta y$ | $y$-distance traveled | EF | Random, in [$-\Delta x$, $\Delta x$] |

**S1 Table. List of parameters used in the model**, including whether it represents an environmental factor (EF), a bacterial trait (BT) or an interaction between them.

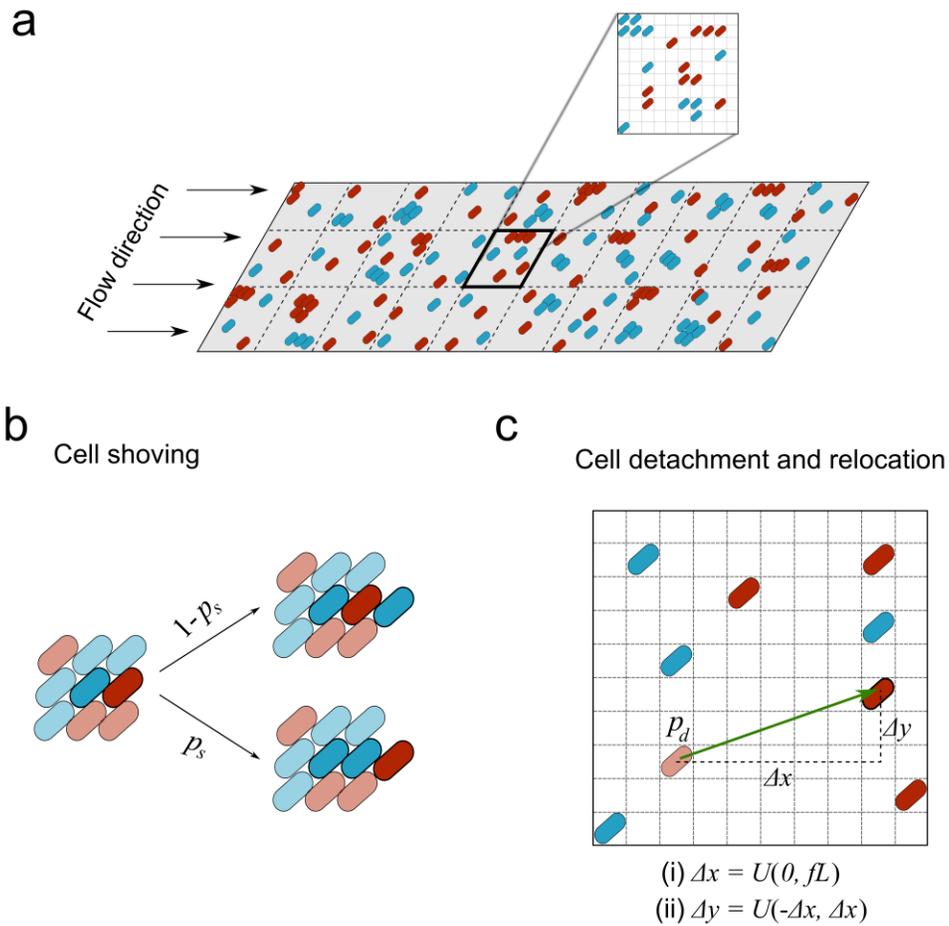

**S1 Fig. Schematic of the experimental setup and the model updating rules.** a) Schematic of the division of the experimental chamber in tiles and model representation of one of the tiles, as a 2D lattice with one cell at each lattice box. b) Cell displacement due to shoving following cell division occurs with probability $p_s$. With complementary probability $1-p_s$ the resident cell keeps its position and the newborn jumps to one of the adjacent empty position. c) Cells may be detached from the surface of the chamber with probability $p_d$ and transported to a new emplacement following the relocation rules explained in the text with periodic boundary conditions (Materials and Methods). $U(a,b)$ indicates a uniformly distributed random variable between $a$ and $b$. $f$ is the flow intensity and $L$ the lattice lateral length.

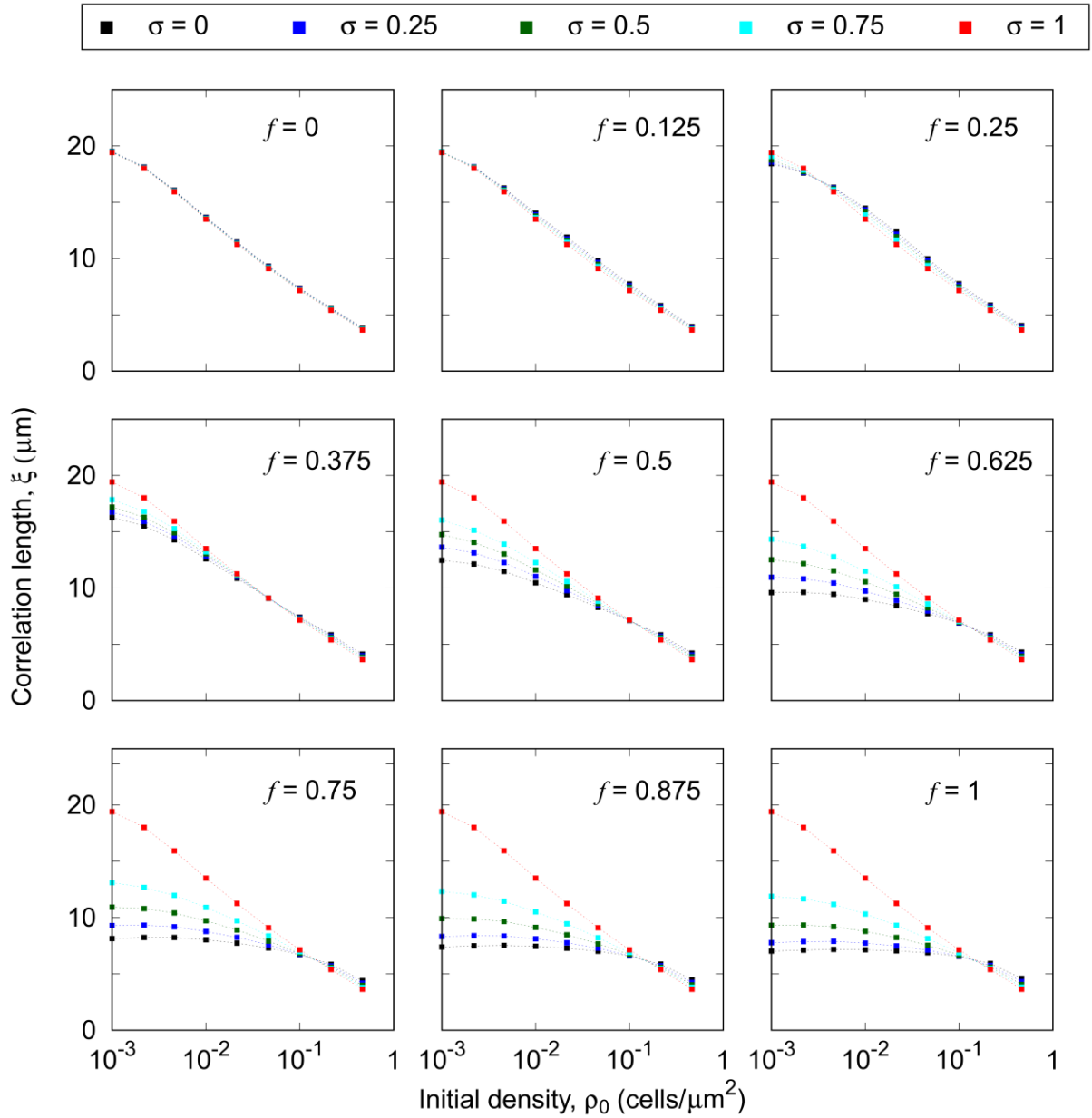

**S2 Fig. Correlation length versus initial density.** Mean correlation length, $\xi$, for different colonization strategies ($\sigma$, $\rho_0$) in several ecological conditions given by the flow intensity $f$. Each curve represents a cell adhesiveness $\sigma$. The color code is maintained in all the panels. Averages are taken over $2\times10^6$ independent model realizations.

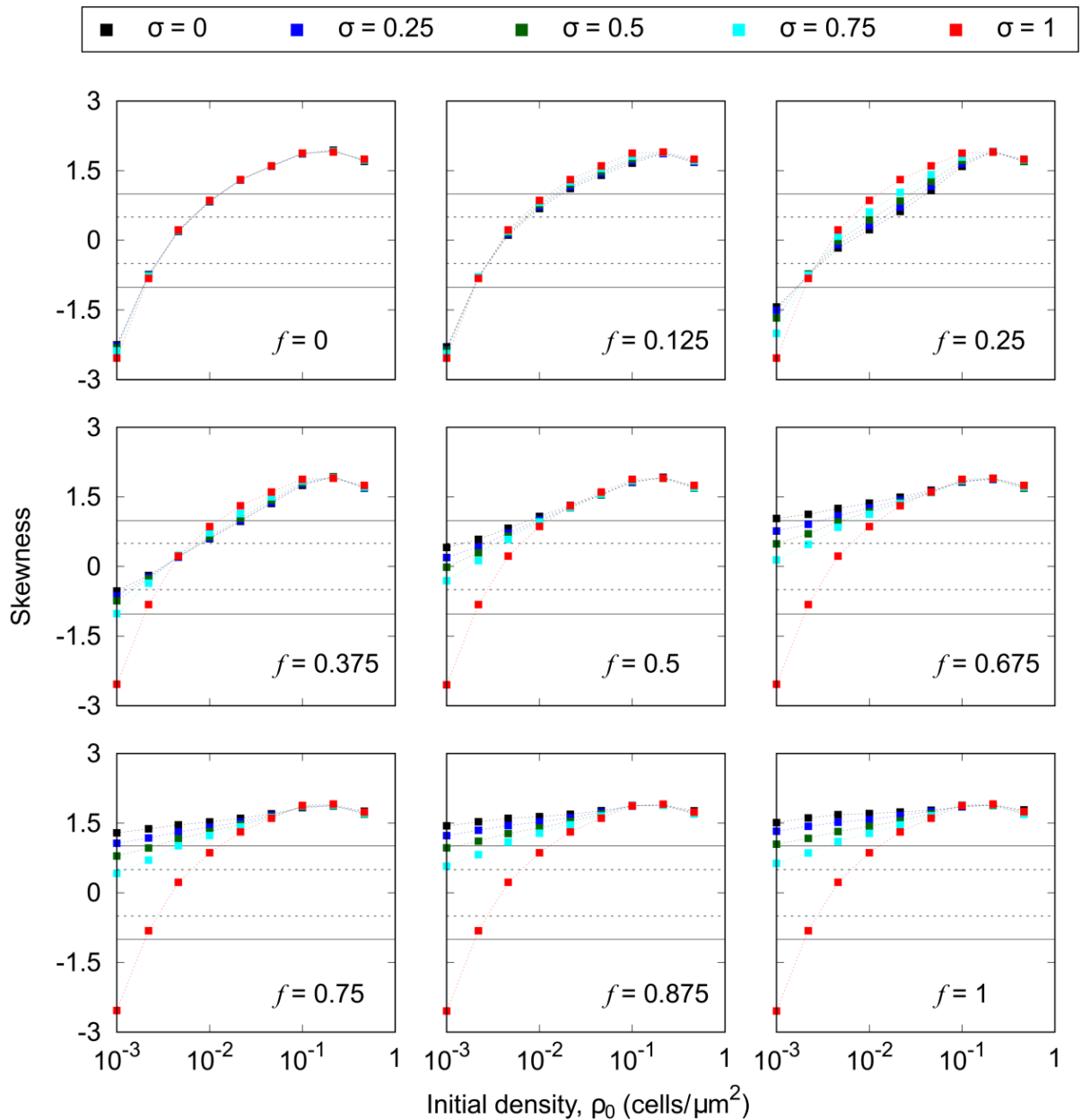

**S3 Fig. Skewness of the correlation length.** Skewness of the distribution of correlation lengths for different colonization strategies ($\sigma$, $\rho_0$) and ecological conditions, given by the flow intensity $f$. Each curve represents a value of the adhesiveness $\sigma$, whose color code is maintained in all the panels. The skewness is obtained from $2\times10^6$ independent realizations of the model. Horizontal dashed lines in each panel indicate the values +/- 0.5 and the full lines, +/- 1. Skewness in the interval [0.5, 1] in absolute value indicate that the data are moderately skewed, and if the skewness greater 1 in absolute value, then the distribution is highly skewed.

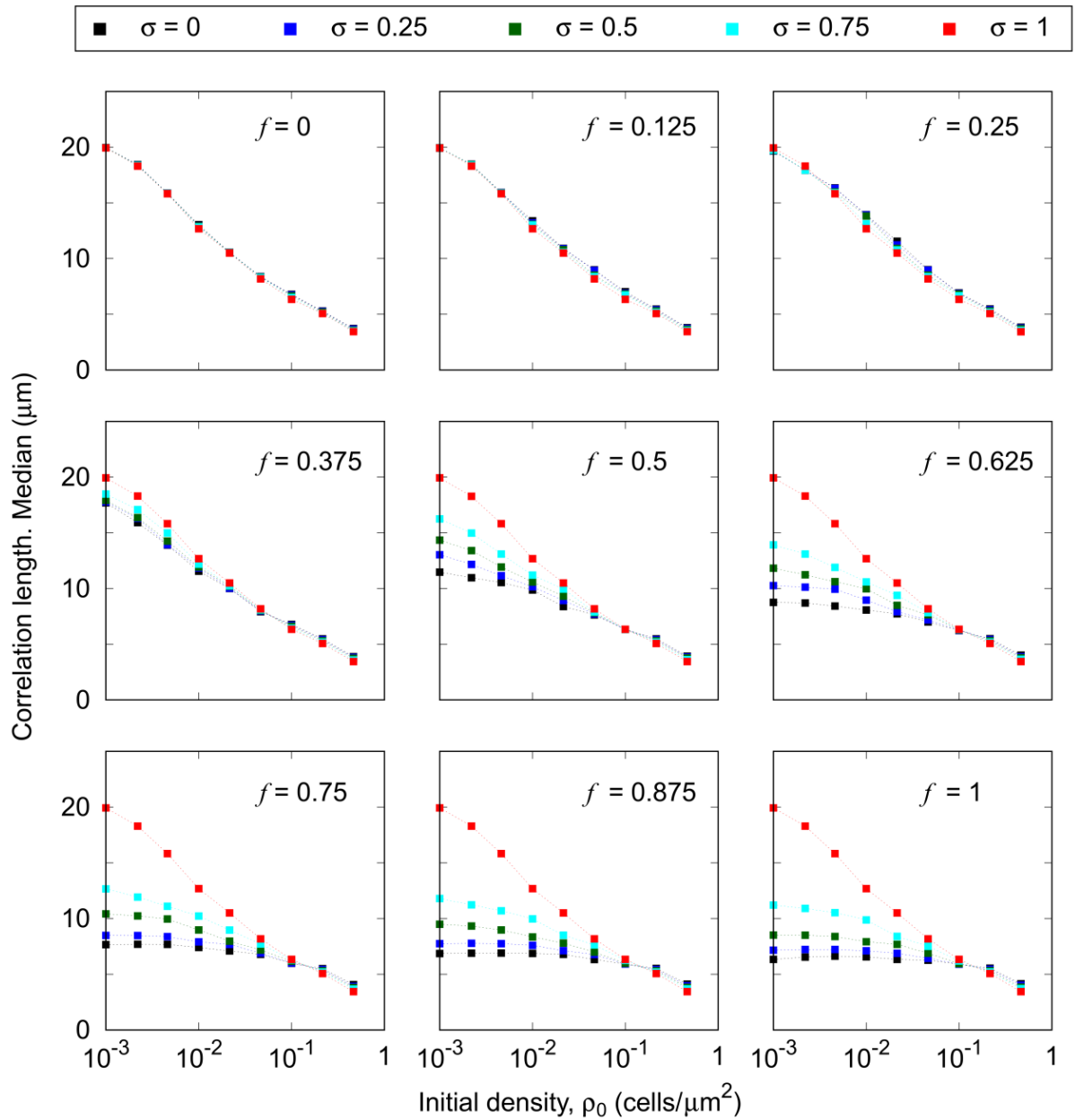

**S4 Fig. Median correlation length.** Median of the correlation length distribution for different colonization strategies ($\sigma$, $\rho_0$) and ecological conditions given by the flow intensity $f$. Each curve represents a value of the adhesiveness $\sigma$. The color code is maintained in all the panels. The median is obtained from a set of $2\times10^6$ independent model realizations.

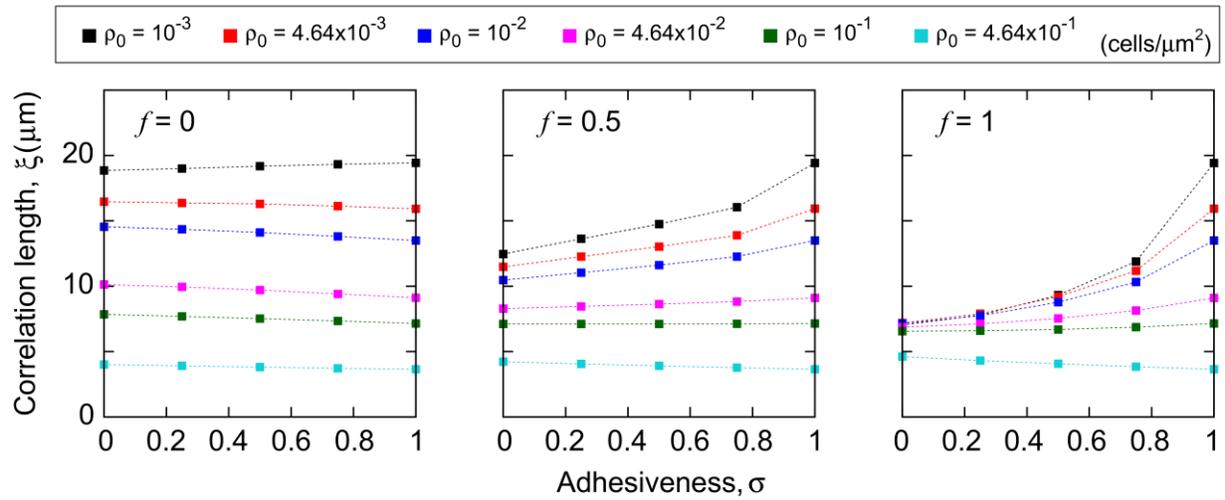

**S5 Fig. Correlation length versus cell adhesiveness.** Mean correlation length, $\xi$, for different colonization strategies ($\sigma$, $\rho_0$) in several ecological conditions given by the flow intensity $f$. Each curve represents a value of the initial density, $\rho_0$. The color code is maintained in all the panels. Averages are taken over $2\times10^6$ independent model realization.

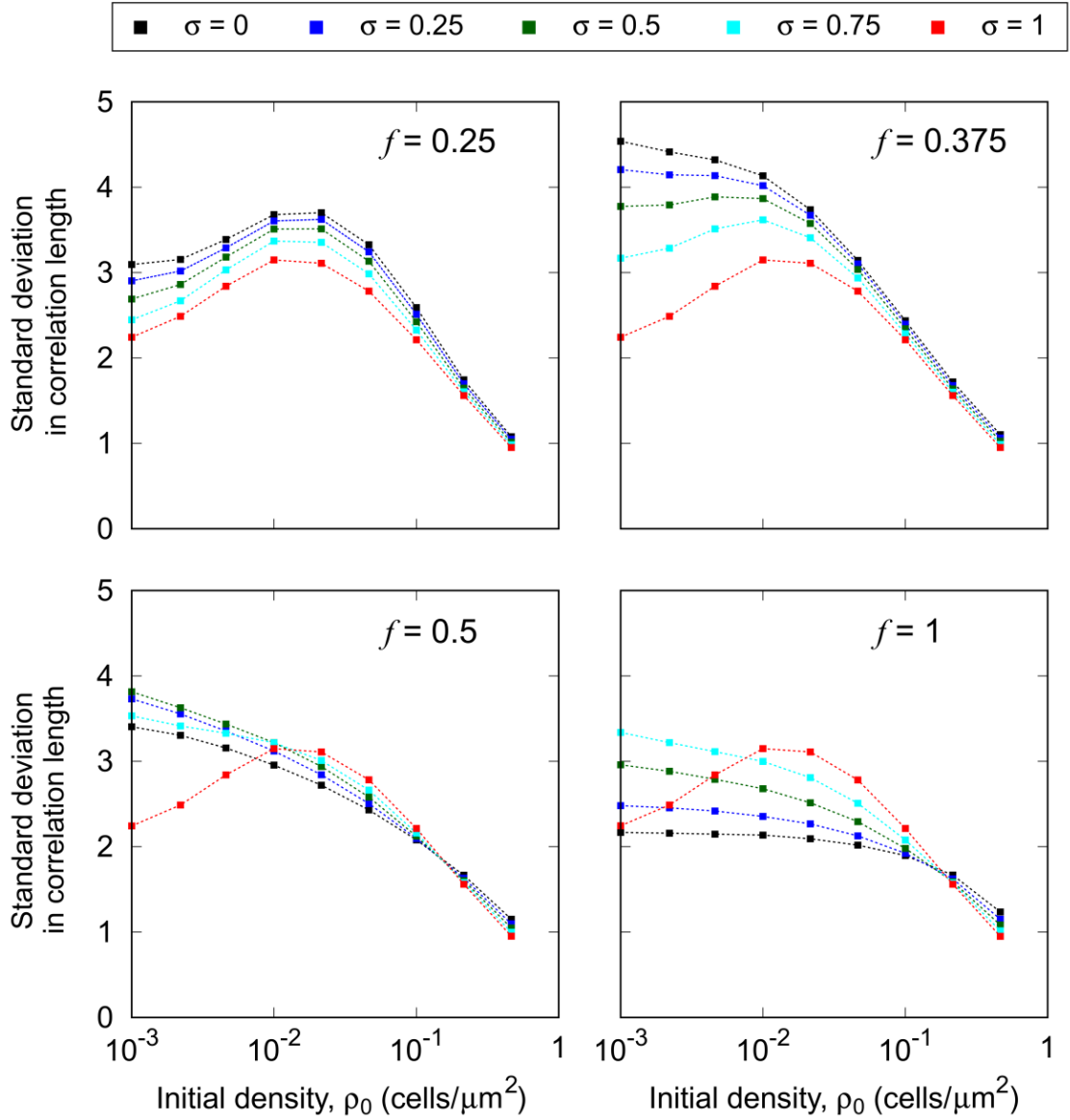

**S6 Fig. Cluster size variability.** a) $f = 0.25$, b) $f = 0.375$, c) $f = 0.5$, d) $f = 1$. Each curve represents the standard deviation in $\xi$ for a given adhesiveness, $\sigma$. Color code is maintained in all the panels. Averages are taken over $2 \times 10^6$ independent model realizations.

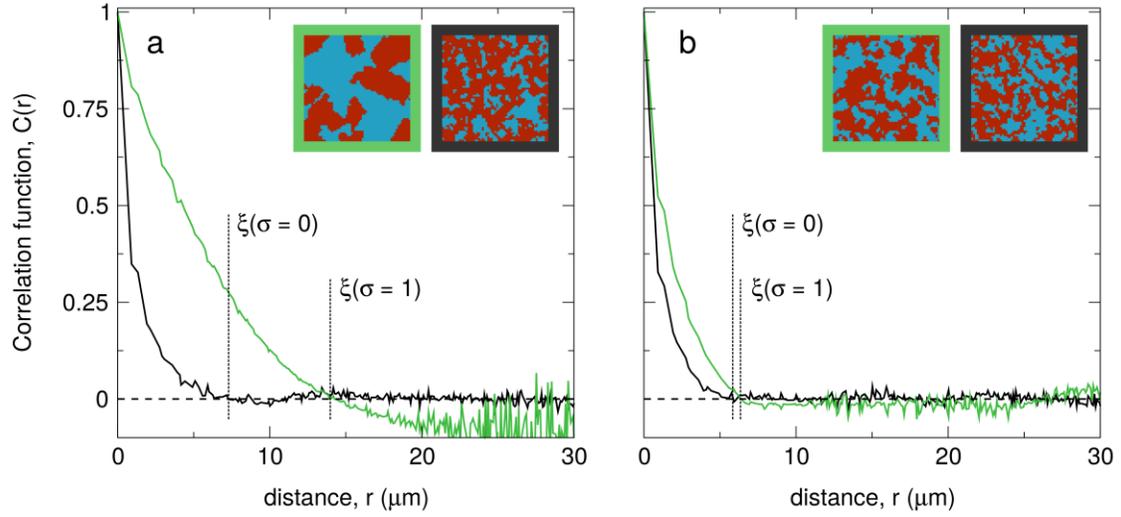

**S7 Fig. Correlation function of individual model realizations.** Correlation functions obtained for single realizations of the model at low (panel a; $\rho_0 = 10^{-3}$ cells/μm$^2$) and high (panel b; $\rho_0 = 10^{-1}$ cells/μm$^2$) initial density of cells. Correlation functions are obtained for the patterns shown in the snapshots. The color code indicates whether the pattern corresponds to $\sigma=1$ (green) or $\sigma=0$ (black) strains. The dashed lines point the value of the correlation length in each case, defined as the first zero of the correlation function.

# S1 Text. Size effect analysis.

Our experimental results have been obtained using a square observation window of lateral length $L = 60\mu m$ embedded within a much larger microfluidic device. Using the simulation framework, we investigated whether the spatial measures in the occupation patterns are influenced by the size of the focal system.

First, we focused on the correlation length, for both highly-adhesive and weakly-adhesive strains, in the intense flow limit ($f = 1$). As shown in the main text, in this regime the model accurately reproduces experimental results if the same focal area is used in both approaches. Numerical simulations on larger systems confirm that both strains maintain the same qualitative trends across simulated areas, and although the curves are quantitatively affected by the simulated area, they intersect at the same value of the initial population density (Fig A1). The sublinear scaling of the correlation length with system size, suggests a saturation of the correlation length in the limit in which $\xi << L$ for any initial density and cell adhesiveness (Fig A2). Next, we prepared a simulation setup in which we divided a system of lateral length $L = 120\mu m$ in four tiles of lateral size $60\mu m$, and simultaneously measured the correlation length in the total system and in each of the tiles. To ensure that the initial population density was constant for the whole system and each tile, we initialized every tile with a total population density $\rho_0$ ($\rho_0/2$ of each strain on average). Focusing on the intense flow limit ($f = 1$), the distance traveled by relocated cells in the direction of the flow is a random number between 0 and $L$, so for a given focal area, the population mixing depends on whether the system is isolated or embedded in a bigger one. However, the use of periodic boundary conditions, as discussed in the main text, minimizes differences in the correlation length for strong flows (Fig A3). The residual difference in the correlation length is due to the fact that, in small isolated systems, the periodic boundary conditions can introduce small additional correlations, since detached cells that exit the system through one of the borders and re-enter through the opposite may be relocated close to their original position. These events are equivalent to limited dispersal and hence tend to increase clonal cluster size. However, as it is shown in Fig A3, their effect is negligible, reinforcing the validity of our periodic boundary conditions.

Next, we extended our analysis to consider a $L=240 \mu m$ patch with various flow intensities. In this scenario, system size influences the outcome of the simulations in two directions. First, the set of flow strengths for which patterns of weakly-adhesive cells have larger clonal clusters than those made by adhesive strains increases considerably. Second, such regions show a larger difference correlation length for bigger systems (Fig A4a). This result indicates that avoiding the production of adhesion substances does not entail a residual gain (slightly larger clusters without the metabolic cost of matrix production) but, for a wide range of environmental flows ($f < 0.4$), such gain can be very significant, as much as that of matrix-production in strong environmental flows (but, again, without the metabolic cost). If, on the other hand, we are observing a small system that, instead of in isolation, is within a bigger one, the flow range for which weakly-adhesive cells show larger clusters segregation is reduced to very weak intensities. This shrinkage of the region results from our flow strength implementation discussed above: when the small system is part of a bigger one, detached cells can travel larger distances even at weak environmental flows and thus the range of limited dispersal is reduced (Fig A4b compared to Figure 4). All this phenomenology indicates that, in a real system, the ratio between the typical distance travelled with the flow and the system size will influence considerably the quantitative (but not the qualitative) behavior of our measure for genetic segregation.

Finally, we analyzed the effect of the system size on cluster size variability for $\sigma = 1$ strains. The standard deviation of the correlation length maintained its concavity regardless of the system size, but it reached its maximum at different initial population densities (Fig A5a). Since highly-adhesive cells are not relocated by the flow, the confluence pattern is strongly determined by the spatial distribution of the founder population, and its correlation length variability depends on the variability of the initial lineage mixing. Hence, it is the number of cells and not the density what determines the position of the maximum in the standard deviation (Fig A5b). For high cell numbers, it is very unlikely to randomly create a configuration with large clusters, whereas for low cell numbers, the cluster size at confluence is necessarily large. In addition, for a fixed initial density (or number of cells) the standard deviation increases with system size since the variability in the spatial distribution of the founder population increases with system size.

In summary, the observation window can quantitatively affect some results of our analyses as well as the regions of the parameter space in which they are expected. Therefore, not only environmental forces, such as fluid flow, and bacterial traits, such as cell adhesion, are important to quantify biofilm population structure. The size of the observation frames needs to be accounted for as well. Importantly, however, the overall qualitative behavior of our results is not affected by the size of the observation window and, therefore, any conclusion drawn for smaller surfaces can be extrapolated to larger systems.

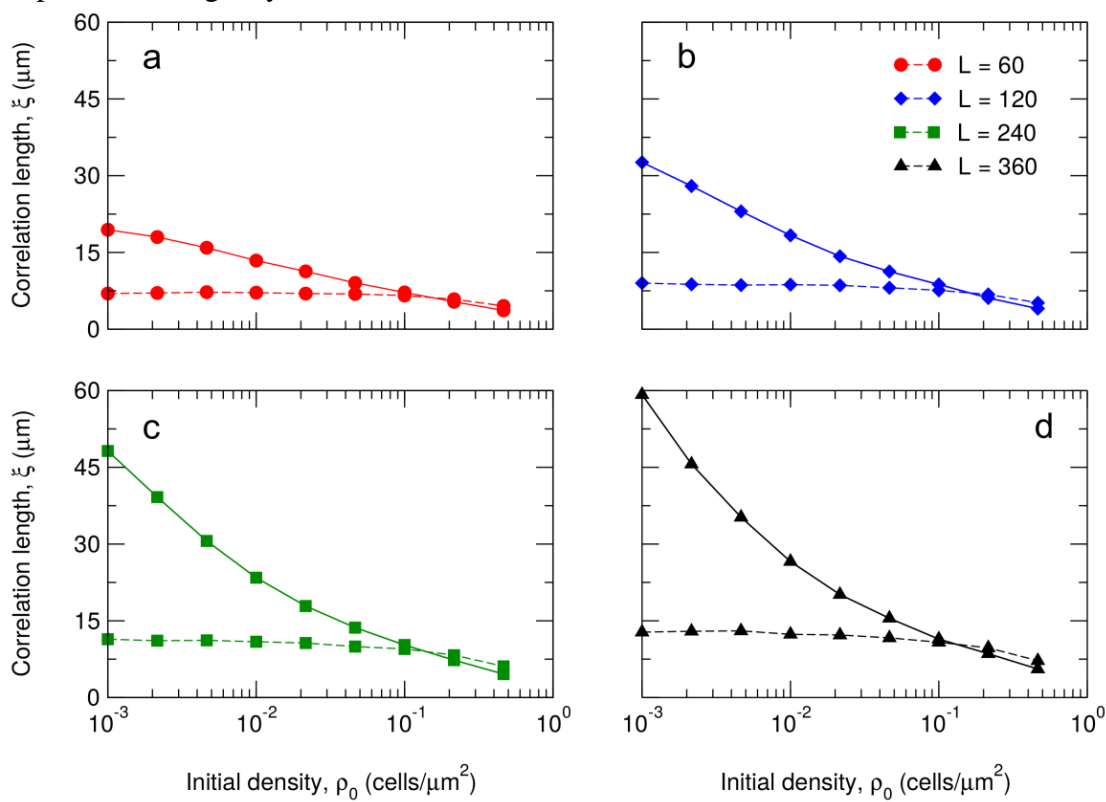

**Fig A1. Effect of surface in correlation length.** The clonal cluster size is strongly influenced by the extension of the colonized surface, although the trends of highly-adhesive and weakly-adhesive strains, and the crossing point between curves, are system size independent. Full lines correspond to $\sigma = 1$ and dashed lines to $\sigma = 0$. a) $L = 60\mu m$, b) $L = 120\mu m$, c) $L = 240\mu m$, d) $L = 360\mu m$.

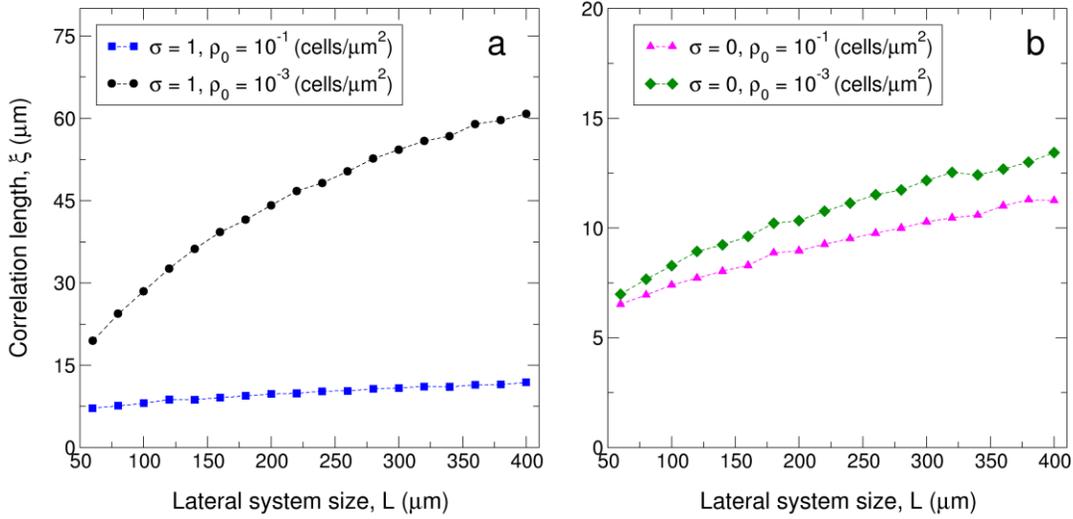

**Fig A2. Correlation length scaling with system size.** The correlation length scales sub-linearly, both for highly-adhesive (a) and weakly-adhesive (b) strains, with system size, which suggests a clonal cluster size saturation for large systems.

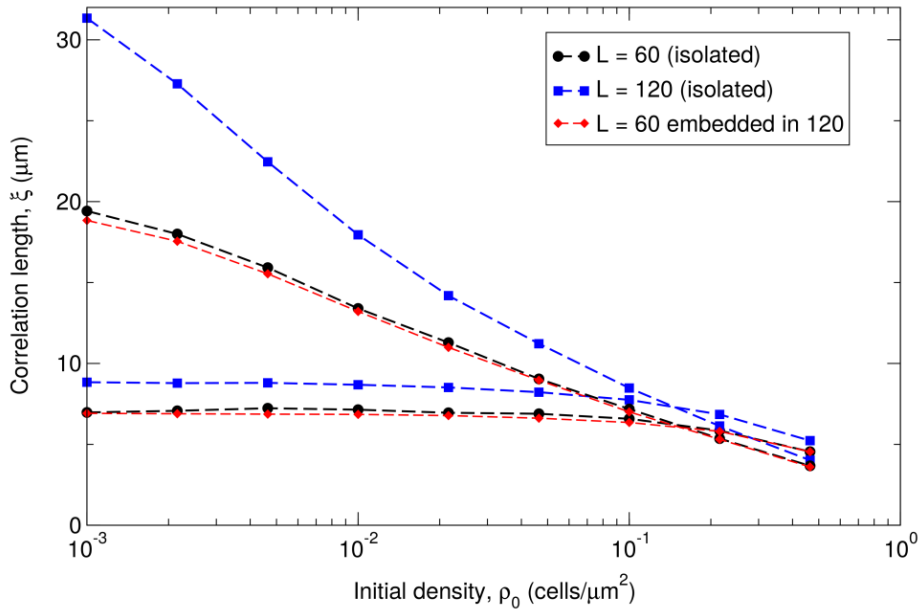

**Fig A3. Finite size effects in the correlation length.** Tiles within a larger system have the same correlation length than isolated surfaces of the same size. Simulations are run independently on systems of lateral length $L = 60\mu m$ (black circles) and $L = 120\mu m$ (blue squares). In this latter scenario, the system is divided in four tiles of lateral length $60\mu m$ and the correlation length of each of the tiles is independently obtained following the same protocol used in the $L = 60\mu m$ case.

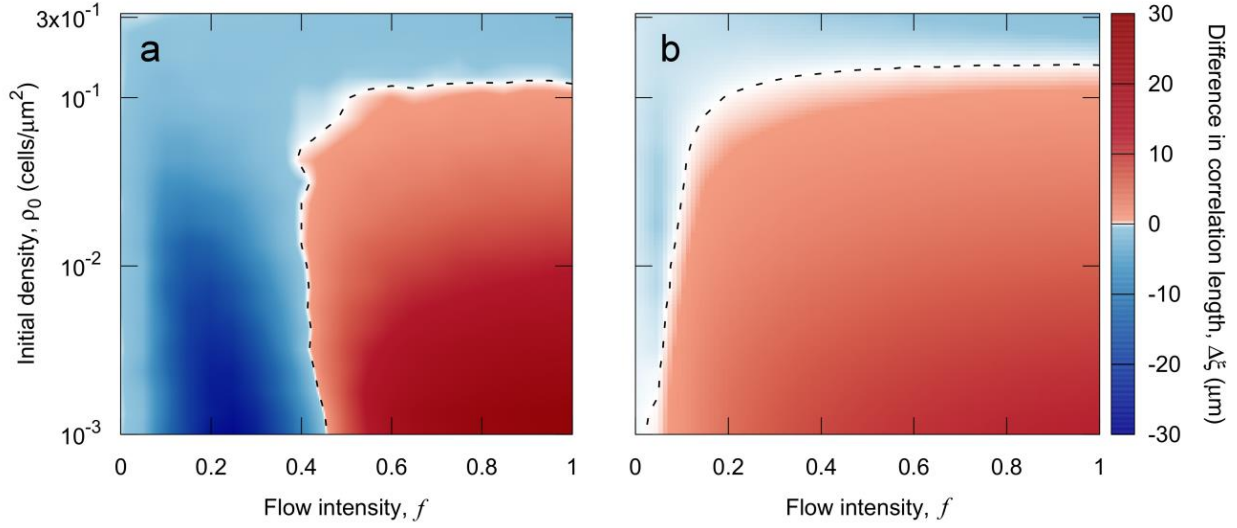

**Fig A4. System size effects on the correlation length difference between highly-adhesive and weakly-adhesive strains.** Correlation length differences are evaluated on a system of lateral length $L = 240 \mu m$ (b) and the result is compared to what would be observed using an observation window of lateral length $L = 60 \mu m$ within the system (a). In the latter case, each of the 16 observation windows is used as an independent replicate. Therefore, averages in b) are taken over 4000 replicates whereas 64000 independent realizations are gathered for the smaller system.

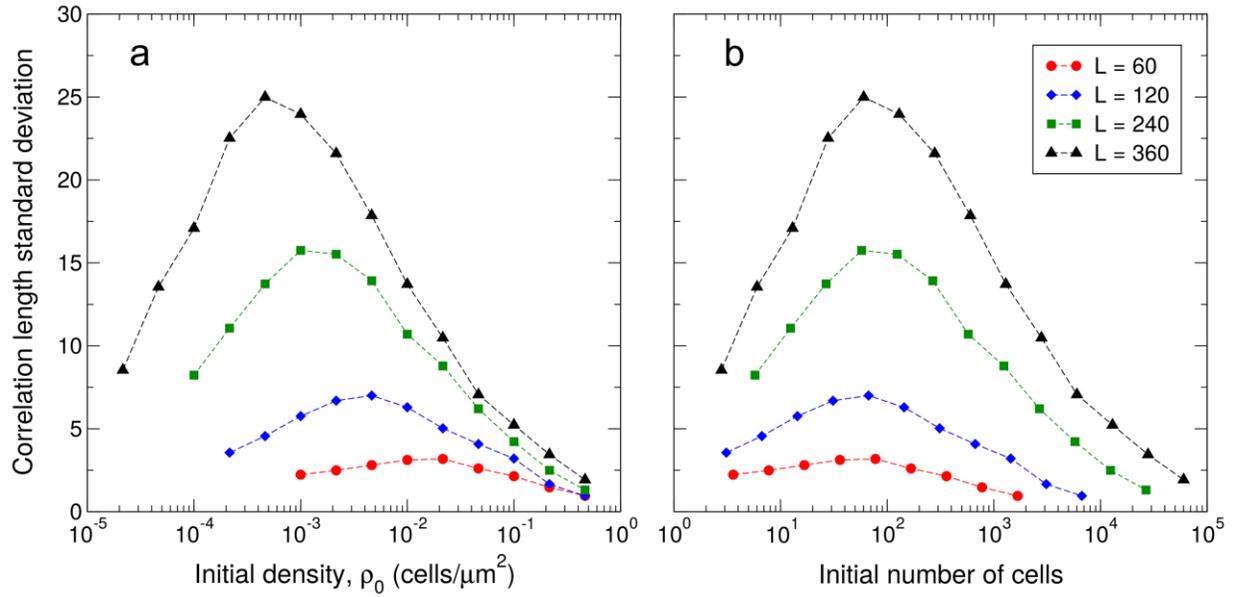

**Fig A5. Variability in the correlation length is influenced by system size.** Correlation length standard deviation versus initial population density (a), and initial number of cells (b). Color code is maintained in both panels